\documentstyle[12pt]{article}
\begin{document}

\title
{\Large \bf Regge Calculus as a Fourth Order
Method in Numerical Relativity}
\author{\bf Mark A. Miller \\
\it \small 201 Physics Building \\
\it \small Syracuse University \\
\it \small Syracuse, New York 13244-1130}
\date{\small (February 27, 1995)}
\maketitle

\begin{abstract}
The convergence properties of numerical Regge calculus as an approximation
to continuum vacuum General Relativity is studied, both analytically
and numerically.  The Regge equations are evaluated on continuum
spacetimes by assigning squared geodesic distances in the continuum
manifold to the squared edge lengths in the simplicial manifold.  It
is found analytically that, individually,
the Regge equations converge to zero as the
second power of the lattice spacing, but that an average over local
Regge equations converges to zero as (at the very least) the third power
of the lattice spacing.  Numerical studies using analytic solutions to the
Einstein equations show that these averages actually converge to
zero as the fourth power of the lattice spacing.
\end{abstract}

\section{Introduction}

In 1961, Regge introduced a discrete form of General Relativity (which has
since been named Regge calculus) \cite{Regge}.
Like General Relativity, Regge calculus
is a geometrical theory of gravity.  However, instead of using
infinitesimal distances (the metric) as the dynamical variables of the
theory, Regge calculus employs {\it finite} distances as dynamical
variables.  Since its introduction, Regge calculus has been
used for a variety of purposes (see \cite{reggereview}), but was
initially envisioned as a tool for numerical relativity \cite{Regge}.
Although work has been done in numerical relativity using Regge calculus,
most applications
share at least one of the two following
undesirable (from the standpoint of practical numerical relativity)
features:
\begin{enumerate}
\item The method used is based on non-simplicial (non-rigid) blocks
(i.e. prisms or hypercubes).
\item The method uses blocks with
Euclidean signature.
\end{enumerate}

The first feature is undesirable for two reasons.  First, not only must
one specify extra ``rigidifying'' conditions on each block, but one must
also make sure that the geometry on a face of a block matches up with
the geometry of the face of an adjacent block (these ``rigidifying''
conditions have non-local constraints).
Simplices, on the
other hand, are rigid: fixing the values of the squared edge lengths
completely determines the geometry inside the simplex, so there
is no problem in trying to match up the geometry of two adjacent
simplices.
Second, it has
recently been realized that Sorkin triangulations (triangulations in
which the Regge equations decouple into single vertex evolutions) can
be constructed from virtually any initial 3-dimensional simplicial manifold,
thus forming a triangulation of the
4-dimensional spacetime \cite{Sorkin,Tuckey,Barrettetal}.

The second feature (Euclidean spacetime) is undesirable since the goal of
numerical relativity is to model physically realistic processes. The
fact is that once one
has a Euclidean implementation of numerical
Regge calculus, it is not an entirely
trivial task to go over to Minkowskian signature.\footnote{This is
a technical point.  Minkowskian signature introduces a bit of added complexity
that must be dealt with, and it may not be as simple as adding a C++ tag
on timelike legs
to an existing Euclidean code, as suggested in
\cite{Galassithesis}.}
When taking derivatives of square roots of negative
functions (as one does in Regge calculus with a $-+++$ signature), one
gets sign flips due to branch cuts in the square root function.  One has
to be very careful to avoid errors that are typified by the
following ``equation'':
\begin{displaymath}
i = \sqrt{-1} = \sqrt{\frac {1}{-1}} = \frac {1}{\sqrt{-1}} = \frac {1}{i}
  = -i
\end{displaymath}

With a goal of using Regge calculus in numerical relativity, I have
developed a code that calculates the Regge equations (and their derivatives
with respect to the squared length of any link) on a simplicial
manifold of arbitrary topology, with each simplex having a flat, Minkowskian
metric.  As a test for the code, I pose the following question:
What is the order of the Regge calculus numerical method?  That is, how
fast do the Regge equations, when ``evaluated'' on a generic solution to
the vacuum Einstein equations, approach zero as the lattice spacing
approaches zero? (How one goes about ``evaluating'' the discrete Regge
equations on a continuous solution of the Einstein equations is described
in the next section.)

In contrast to Regge calculus, the order of a given finite differencing
scheme is clear from the beginning: it is the order of the error introduced
by truncating the Taylor expansion when approximating functions and their
derivatives on a discrete lattice.  For example, the derivative of
a function $f(x)$ defined on grid points $x_i$ (evenly spaced; $x_{i+1}
- x_i = h$) can be approximated at $x_0$ with a Taylor expansion about
$x = x_0$:
\begin{displaymath}
f(x_1) = f(x_0 + h) = f(x_0) + f^{\prime}(x_0) h + {\cal{O}} (h^2)
\end{displaymath}
namely,
\begin{displaymath}
f^{\prime}(x_0) = \frac {f(x_1) - f(x_0)}{h} + {\cal{O}} (h)
\end{displaymath}
This is a {\it first order} approximation to the derivative of $f(x)$ at
$x_0$.  In a similar manner, any differential operator, ${\cal{L}}$, acting
on a set of functions, $\vec{f}(\vec{x})$, can be approximated by an
$n^{th}$ order finite difference operator, $\hat{\cal{L}}_{n}$, acting
on functions $\vec{f}$ defined on a lattice $\vec{x}_i$ characterized
by uniform lattice spacing h.  If $\vec{f_0} (\vec{x})$ is a solution
to the differential equation
\begin{displaymath}
{\cal{L}} {\vec{f}} = 0
\end{displaymath}
then, by construction, the finite difference operator  $\hat{\cal{L}}_{n}$
acting on $\vec{f_0} (\vec{x_i})$ (the continuum solution $\vec{f_0}(\vec{x})$
evaluated at lattice sites $\vec{x_i}$) will converge to zero as
the $n^{th}$ power of h:
\begin{displaymath}
\hat{{\cal{L}}_n} \vec{f_0} (\vec{x_i}) = 0 + p_0 h^n + {\cal{O}}(h^{n+1})
\end{displaymath}
where $p_0$ is some constant.

In the case of Regge calculus, it is not known, a priori, what order (in
lattice spacing h) of error is introduced by discretizing the Einstein
equations.  This question is the subject of this paper.  I will show
analytically that, individually, the Regge equations, when evaluated
on a continuum spacetime, converge to zero as the second power of the
lattice spacing (there is a non-zero second order term).\footnote{This
fact has been independently observed by Leo Brewin\cite{Brewinpreprint}}
This result will be seen to be independent of whether or
not the continuum metric satisfies the Einstein equations.
However, it will be
shown analytically that averages (with weightings given by variations
of each component of the continuum metric) over local Regge equations
have a vanishing second order
 term only when the continuum metric satisfies the
Einstein equations.
Studying the behavior of the
third order term must be done numerically, using
analytic solutions to the Einstein equations.  These results show that,
for the specific solutions studied,
the third order term of the average of the Regge equations also vanishes.

\section{Regge Calculus}

Regge calculus is based on the idea of approximating a curved continuum
spacetime with a collection of flat simplices.  For instance, a curved
2-dimensional surface can be approximated by a collection
of 2-simplices (triangles).
A general property of simplicial manifolds is that the curvature always
resides on subspaces of codimension two.  In the 2-dimensional case,
this means that the curvature is concentrated on the vertices.  In fact,
the integral of the scalar curvature over a region of the simplicial
manifold is simply a sum of the deficit angles associated with each
vertex.

In four dimensions, the curvature is concentrated on 2-dimensional
simplices (triangles).
The Regge action is simply the Hilbert action evaluated on the
simplicial manifold:
\begin{equation}
S(l^2) = \int \sqrt{-g} R = \sum_{i} A_i {\epsilon}_i
\end{equation}
where the sum on the right hand side is over all triangles (labeled by
$i$), $A_i$ is the area of triangle $i$, and ${\epsilon}_i$ is the
deficit angle associated with the $i^{th}$ triangle
(boundary terms in the action are treated in \cite{hartlesorkin}).  Here, we
have set $16 \pi G = 1$.
Demanding that the action be stationary with respect to small variations in
the squared edge lengths yields the Regge equations.  Each dynamical
variable (squared edge length) has an associated Regge equation.  The
Regge equation associated with leg $(i,j)$ is
\begin{equation}
{\cal R}_{ij}(l^2) = \sum_k {\epsilon}_{ijk} \cot{ {\theta}_{ijk}} = 0
\end{equation}
where the sum is over all triangles $(i,j,k)$ that have $(i,j)$ as one
edge, ${\theta}_{ijk}$ is the angle $\angle ikj$ of triangle $(i,j,k)$,
and ${\epsilon}_{ijk}$ is the deficit angle associated with triangle
$(i,j,k)$.

One property of the Regge equations that will be important in the following
sections is the fact that the equations are dimensionless.  Only angles
appear in the Regge equations.  Specifically, look at the expression
for $\cot{{\theta}_{ijk}}$ in terms of the squared edge lengths
$l^2_{ij}$, $l^2_{ik}$, and $l^2_{jk}$:
\begin{equation}
\cot{{\theta}_{ijk}} =
\frac {l^2_{ik} + l^2_{jk} - l^2_{ij}}{\sqrt{2 l^2_{ij} l^2_{ik} +
2 l^2_{ij} l^2_{jk} + 2 l^2_{ik} l^2_{jk} - l^4_{ij} - l^4_{ik}
- l^4_{jk} }}
\end{equation}
We see explicitly that both the numerator and denominator have units of
$L^2$.  In section \ref{sec-analytic},
I will want to substitute a power series
(in smallness parameter $\delta$) for each $l^2$ which has the form
\begin{equation}
l^2_{ij} = {(l^2_2)}_{ij} {\delta}^2 + {(l^2_3)}_{ij} {\delta}^3 +
           {(l^2_4)}_{ij} {\delta}^4 + \cdots  \label{eq:lsquared}
\end{equation}
Observe that $\cot{{\theta}_{ijk}}$ (and, therefore, the Regge equation)
would be indeterminate $(0/0)$ when $\delta = 0$.  In doing the actual
calculations, equation (\ref{eq:lsquared}) is scaled by a factor
of $1/{{\delta}^2}$.  The first non-zero term in a typical
series expansion for $l^2$ will now be of order unity, and the
Regge equations will be perfectly well defined at $\delta = 0$. Due to the
complexity of the Regge equations, MATHEMATICA was used to perform the analytic
calculations in section (\ref{sec-analytic}).

\section{A Formalism for Studying the Convergence Properties
of the Regge Equations}
\label{sec-lattice}

In the past, various methods have been used to study the
relationship between Einstein's General Relativity in the continuum
and the discrete Regge calculus.  Both Sorkin \cite{Sorkin} and
Friedberg and Lee \cite{FriedbergandLee} derive the Regge action from
the Hilbert action by considering the limit of a sequence of continuum
surfaces that approximate a piecewise flat surface.  Barrett
\cite{Barrett1987a} showed the equivalence between the Regge equations
and the vanishing of the flow of energy-momentum across a certain
hypersurface.
Brewin \cite{Brewin} has shown
the equivalence of the Regge action and the Hilbert action on
almost flat simplicial spacetimes by treating the Riemann tensor as
a distribution.

The basic idea I use for comparing the continuum Einstein equations to the
Regge equations is the same as used in a paper
by Cheeger {\it et al} \cite{Cheegeretal} (this idea can also
be found in \cite{Sorkin}): assign the
(signed) square of the geodesic length
between two points in the continuum manifold whose
metric satisfies the vacuum Einstein equations to the
squared edge
length of the corresponding link in the simplicial manifold.  In
\cite{Cheegeretal}, it was shown (for Euclidean metrics)
that the Regge action converges to
the Hilbert action, in the sense of measures, as long as certain
fatness conditions are satisfied by the simplices.  Here, I ask a
different question: how {\it fast} do the Regge equations converge to
zero as the lattice spacing goes to zero?

\subsection{Evaluating the Regge Equations on a Continuum Spacetime}

\begin{sloppypar}
For simplicity, this paper will use a hypercubic
simplicial topology (hypercubes split into simplices).  Consider
a 4-manifold, ${\cal M}$, equipped with a metric ($g_{\mu \nu}$) whose
signature is everywhere ($-+++$).  Furthermore, let $g_{\mu \nu}$ be
a solution to the vacuum Einstein equations $R_{\mu \nu} = 0$.  Let
$\{ x^\mu \}$ be coordinates on a region of ${\cal M}$ containing
some point ${\cal O}$ with coordinates $x^{\mu}_{\cal O} = (0,0,0,0)$.
Now, introduce the hypercubic simplicial triangulation in the following
way.  First, construct a rectangular lattice in the coordinates $\{ x^\mu \}$.
The lattice site $x_{\vec{n}}$ (where $\vec{n} = (n^0,n^1,n^2,n^3)$) is
located at coordinates $x^{\mu}_{\vec{n}} = (n^0 \; \Delta x^0,
n^1 \; \Delta x^1, n^2 \; \Delta x^2, n^3 \; \Delta x^3)$, where
$\Delta x^\mu$ are 4 arbitrary lattice spacing constants of order unity.
To construct a triangulation from this rectangular lattice, we
construct 15 links at each lattice site $x_{\vec{n}}$:
\begin{displaymath}
\begin{array}{cc}
\verb+diagonal links+ & \verb+rectangular links+  \\
(x_{\vec{n}}, x_{\vec{n} + \hat{n}_0 + \hat{n}_1}) &  \\
(x_{\vec{n}}, x_{\vec{n} + \hat{n}_0 + \hat{n}_2}) &  \\
(x_{\vec{n}}, x_{\vec{n} + \hat{n}_0 + \hat{n}_3}) &  \\
(x_{\vec{n}}, x_{\vec{n} + \hat{n}_1 + \hat{n}_2}) &
(x_{\vec{n}},x_{\vec{n} + \hat{n}_0}) \\
(x_{\vec{n}}, x_{\vec{n} + \hat{n}_1 + \hat{n}_3}) &
(x_{\vec{n}},x_{\vec{n} + \hat{n}_1}) \\
(x_{\vec{n}}, x_{\vec{n} + \hat{n}_2 + \hat{n}_3}) &
(x_{\vec{n}},x_{\vec{n} + \hat{n}_2}) \\
(x_{\vec{n}}, x_{\vec{n} + \hat{n}_0 + \hat{n}_1 + \hat{n}_2}) &
(x_{\vec{n}},x_{\vec{n} + \hat{n}_3}) \\
(x_{\vec{n}}, x_{\vec{n} + \hat{n}_0 + \hat{n}_1 + \hat{n}_3}) & \\
(x_{\vec{n}}, x_{\vec{n} + \hat{n}_0 + \hat{n}_2 + \hat{n}_3}) & \\
(x_{\vec{n}}, x_{\vec{n} + \hat{n}_1 + \hat{n}_2 + \hat{n}_3}) & \\
(x_{\vec{n}}, x_{\vec{n} + \hat{n}_0 + \hat{n}_1 + \hat{n}_2 + \hat{n}_3}) &
\end{array}
\end{displaymath}
In this way, each hypercube is split into 24 4-simplices.
We now have a simplicial manifold with each vertex $x_{\vec{n}}$
associated with a point on $\cal M$ (the point with coordinates
$(n^0 \; \Delta x^0,
n^1 \; \Delta x^1, n^2 \; \Delta x^2, n^3 \; \Delta x^3)$).
In order to study the limit of small lattice spacings, we multiply the
coordinates corresponding to each lattice site with a ``smallness parameter'',
$\delta$.  The coordinates associated with lattice site $x_{\vec{n}}$ are now
$(\delta \; n^0 \; \Delta x^0,
\delta \; n^1 \; \Delta x^1, \delta \; n^2 \; \Delta x^2,
\delta \; n^3 \; \Delta x^3)$.
\end{sloppypar}

We now assign the (signed) square of the geodesic length of separation
between points on the continuum manifold to the squared lengths of the links
in the simplicial manifold (see
Figure 1).  For instance, to find the squared length of the
link ($x_{(0,0,0,0)}, x_{(1,0,0,0)}$) in the simplicial manifold,
\begin{enumerate}
\item find the solution to the geodesic equation,
\begin{displaymath}
\frac {d^2 x^{\mu}(\lambda)} {d {\lambda}^2} + {\Gamma}^{\mu}_{\alpha
\beta}(x(\lambda)) \frac {d x^{\alpha}(\lambda)}{d \lambda}
                   \frac {d x^{\beta}(\lambda)} {d \lambda} = 0
\end{displaymath}
with boundary conditions $x^{\mu}(\lambda = 0) = (0,0,0,0)$ and
                         $x^{\mu}(\lambda = 1) = (\delta \; \Delta x^0,0,0,0)$,
where $\lambda$ is an affine parameter ranging for 0 to 1.
\item integrate the length of the geodesic :
\begin{displaymath}
l = \int_{0}^{1} d \lambda \; \sqrt{g_{\mu \nu} \frac {d x^{\mu}}{d \lambda}
\frac {d x^{\nu}}{d \lambda}}
\end{displaymath}
\item set the squared length of the link
equal to $l^2$ (this quantity will be positive for
spacelike geodesics, and negative for timelike geodesics).
\end{enumerate}

\section{Analytic Results}
\label{sec-analytic}

The geodesic equation cannot be solved analytically for an arbitrary
metric.  However, since all we want to do is find the first non-vanishing
order of the Regge equations, we will be content with a power expansion
(in $\delta$)
for the squared geodesic distance of curves in the continuum manifold.

\subsection{Series Expansion for the Squared Geodesic Distances}
\label{sec-geoseries}

In general, we will need to find the power series expansion in $\delta$
for the squared geodesic distances between two points ${\cal P}_i$ and
${\cal P}_f$ with respective coordinates $\{\delta \; x^{\mu}_i\}$ and
$\{\delta \; x^{\mu}_f\}$ in terms of the metric at $\cal O$, the
Christoffel symbols at $\cal O$, the Riemann tensor at $\cal O$, the
first derivatives of the Riemann tensor at $\cal O$, etc.  Let $x^{\mu}
(\lambda)$ be a parametrization of the geodesic with affine parameter
$\lambda \in [0,\delta]$ (so that
the affine parameter is of the same order as the length of the
geodesic) that satisfies the boundary conditions
\begin{equation}
\begin{array}{c}
x^{\mu}\mbox{\scriptsize ($\lambda = 0$)} = \delta \; x^{\mu}_i    \\
x^{\mu}\mbox{\scriptsize ($\lambda = \delta$)} = \delta \; x^{\mu}_f
\end{array}
\label{eq:boundary}
\end{equation}
Now, expand $x^{\mu}(\lambda)$ about the point $\lambda = 0$:
\begin{equation}
x^{\mu}(\lambda) = x^{\mu}(0) + \frac {d x^{\mu}}{d \lambda}(0) \, \lambda +
\frac {1}{2!} \frac {d^2 x^{\mu}}{d {\lambda}^2}(0) \, {\lambda}^2 +
\frac {1}{3!} \frac {d^3 x^{\mu}}{d {\lambda}^3}(0) \, {\lambda}^3 + \cdots
\label{eq:coordexpand}
\end{equation}
Taking this expansion, along with an expansion of the Christoffel
symbols about $\cal O$ and substituting into the geodesic equation allows
one to solve for the coefficients of eq. (\ref{eq:coordexpand})
in terms of the boundary
values (the coordinates of the endpoints
of the geodesic) and the Christoffel symbols and their
derivatives at point ${\cal O}$.
One then expands the metric as a power series about $\cal O$, and
substitutes it into the final expression,
\begin{equation}
l^2_{i,f} = {\left(\int_0^\delta \, d \lambda \, \sqrt{g_{\mu \nu}
\frac {d x^{\mu}}{d \lambda} \frac {d x^{\nu}}{d \lambda}}\right)}^2 =
{\left( g_{\mu \nu} \frac {d x^{\mu}}{d \lambda} \frac {d x^{\nu}}{d
\lambda} \right)}_{\lambda = 0} {\delta}^2
\end{equation}
along with the power series of $\frac {d x^\mu}{d \lambda}$ (obtained by
rearranging eq. (\ref{eq:coordexpand}))
to get the final power series expansion for
the square of the geodesic distance between points ${\cal P}_i$ and
${\cal P}_f$.

\subsection{Expanding the Regge Equations as a Power Series}

With a power series expansion for the squared geodesic distances in hand, we
now evaluate the Regge equations on the simplicial manifold, where each squared
length in the simplicial manifold is assigned the squared geodesic
distance between the corresponding two points in the continuum manifold.
A typical Regge equation can then be expressed as a power series in $\delta$,
\begin{equation}
{\cal R}_{ij}(\delta) = {\cal R}_{ij}(0) +
\left( \frac {d {\cal R}_{ij}}{d \delta}(0)\right)
\delta +
\frac{1}{2!} \left( \frac {d^2 {\cal R}_{ij}}{d {\delta}^2}(0) \right)
{\delta}^2 + \cdots
\label{eq:reggeexpand}
\end{equation}
where the coefficients of the series are functions of the metric
at point $\cal O$, the Christoffel symbols at point $\cal O$, the
first derivatives of the Christoffel symbols at point $\cal O$, etc.
Recall that, due to the dimensionlessness of the Regge equations, in
order to calculate the $n^{th}$ order coefficient in the expansion of a
particular Regge equation, all of the expressions for the squared geodesic
distances must be accurate up to and including the $(n+2)^{th}$ order in
$\delta$.

\subsubsection{Zeroth Order Results}

To calculate the zeroth order term in the expansion of a typical
Regge equation, eq. (\ref{eq:reggeexpand}),
we require expressions for squared geodesic
distances that are accurate up to and including the second order
in $\delta$.  The analysis
in section \ref{sec-geoseries} gives us the answer:
\begin{equation}
l^2_{if} = {(g_{\mu \nu})}_{\cal O}(x^{\mu}_f - x^{\mu}_i)
(x^{\nu}_f - x^{\nu}_i) {\delta}^2 + {\cal O}({\delta}^3)
\end{equation}
Substituting these expressions into a typical Regge equation, and setting
$\delta = 0$ is equivalent to evaluating the Regge equations on a continuum
manifold with constant metric ${(g_{\mu \nu})}_{\cal O}$.  Therefore, all
of the deficit angles vanish, and the zeroth order term in the expansion
of a typical Regge equation, eq. (\ref{eq:reggeexpand}), is zero.

\subsubsection{First Order Results}

To calculate the first order term in the expansion of a typical Regge
equation, eq. (\ref{eq:reggeexpand}),
we require expressions for squared geodesic
distances that are accurate up to and including the third
order in $\delta$. Again, the
analysis in section \ref{sec-geoseries} gives us the answer:
\begin{equation}
l^2_{if} = {(g_{\mu \nu})}_{\cal O}(x^{\mu}_f - x^{\mu}_i)
(x^{\nu}_f - x^{\nu}_i) {\delta}^2 +
{(\Gamma_{\alpha \mu \nu})}_{\cal O}(x^{\mu}_i + x^{\mu}_f)
(x^{\nu}_f - x^{\nu}_i)(x^{\alpha}_f - x^{\alpha}_i){\delta}^3 +
{\cal O}({\delta}^4)
\label{eq:firstl}
\end{equation}
The first order term in the expansion of a typical Regge equation,
$(\frac {d {\cal R}_{ij}} {d \delta})_{\delta = 0}$, has been
computed on a number of different lattices (hypercubic lattice,
randomly generated lattice, quantity production lattice \cite{qpl}), all
with the same result:
\begin{equation}
{(\frac {d {\cal R}_{ij}}{d \delta})}_{\delta = 0} = 0
\end{equation}
This result can be interpreted geometrically
as a statement of the fact that there
is no first order change in a vector that is parallel transported around
a small closed loop.  From the standpoint of numerical relativity,
this result means that, at the very least, Regge calculus is a
second order numerical method (Of course, the result is independent of
whether or not the metric satisfies the Einstein equations,
since eq. (\ref{eq:firstl})
depends only on the metric and its first derivatives (Christoffels) at
point ${\cal O}$).

\subsubsection{Second Order Results}

Unfortunately, the problem of expressing the squared geodesic distances
(in an arbitrary coordinate system), accurate up to
and including the fourth order in $\delta$ (needed
to study the second order behavior of the Regge equations), in
terms of the Riemann tensor (as well as the metric and Christoffel symbols)
at point ${\cal O}$ seems to be intractable.  To remedy this, I will
use Riemann normal coordinates \cite{MTW} about point ${\cal O}$.  One
property of Riemann normal coordinates is that the Christoffel symbols
vanish at point ${\cal O}$.
 From eq. (\ref{eq:firstl}), it can be seen that this
causes the third order term in the expression for the squared geodesic
distance to vanish.  Using other properties
of Riemann normal coordinates, namely,
\begin{equation}
{(g_{\mu \nu})}_{\cal O} = {\eta}_{\mu \nu}
\end{equation}
\begin{equation}
{( {\partial}_{\alpha} {\Gamma ^ \nu}_{\beta \mu})}_{\cal O} =
- \frac {1}{3} {( {R_{\alpha \mu \beta}}^\nu + {R_{\alpha \beta \mu}}^\nu )}
_{\cal O}
\end{equation}
\begin{equation}
{( {\partial}_{\alpha} {\partial}_{\beta} g_{\mu \nu} )}_{\cal O}
= - \frac {1}{3} {(R_{\alpha \mu \beta \nu} + R_{\beta \mu \alpha \nu} )}
_{\cal O}
\end{equation}
one can obtain the expression for the squared geodesic distance that
is accurate up to and including the
fourth order in $\delta$:
\begin{equation}
l^2_{if} = {({\eta}_{\mu \nu})}(x^{\mu}_f - x^{\mu}_i)
(x^{\nu}_f - x^{\nu}_i) {\delta}^2 -
\frac {1}{3} {(R_{\alpha \mu \beta \nu})}_{\cal O}
x^{\alpha}_i x^{\beta}_f  (x^{\mu}_f - x^{\mu}_i)
(x^{\nu}_f - x^{\nu}_i) {\delta}^4 +
{\cal O}({\delta}^5)
\label{eq:secondl}
\end{equation}

One would hope that when substituting the above expression for squared
geodesic distances into the Regge equation
expansion (\ref{eq:reggeexpand}), one would
observe the vanishing of the second order term when the Einstein
equations are imposed (${(R_{\mu \nu})}_{\cal O} = 0$).  Although
there are cases (particularly when there is high lattice symmetry)
where the second order term does vanish when the Einstein equations
are imposed, it does not vanish in general.

\subsection{Averaging Over a Region of Spacetime}

 From the above analysis, we see that only the zeroth and first order
terms in the expansion of a typical Regge equation vanish.  However, this
result holds even if the continuum metric does {\it not} satisfy the
Einstein equations!!  Therefore, how confident can one be that
Regge calculus is a good approximation to General Relativity?
To motivate an answer to this, consider the analogous problem
of approximating a continuous curve with a series of straight line segments.
If we look closely at any one particular line segment, it does
not look as if we are approximating the continuum curve very well.  However,
if we ``step back'' and look at many line segments,
the approximation is better (see Figure 2).

Therefore, we might expect a suitable average of the Regge equations
contained in some region of spacetime to have a vanishing second order term
when the Einstein equations are imposed on the continuum metric \cite{sorkin1}.
How should one go about constructing this average?  What should one use
as a reasonable weighting?  The continuum action principle suggests a partial
answer\cite{sorkin1}.
If we view the squared leg lengths in the Regge action as
functions of the continuum metric,
\begin{equation}
S = S(l^2_{ij}(g_{\mu \nu}))
\end{equation}
then the variation of the action
with respect to variations in the metric ($\delta g_{\mu \nu}$) is simply
\begin{equation}
\delta S = \sum_{ij} \left( \frac {\partial S}{\partial l^2_{ij}} \right)
\left( \frac {\partial l^2_{ij}}{\partial g_{\mu \nu}} \right)
\; \delta g_{\mu \nu}
\end{equation}
The first factor in the above sum, $ \partial S / \partial l^2_{ij}$, is
the Regge equation that corresponds to leg $(i,j)$.  The
second factor can be used as the weighting of that particular
Regge equation.  There will be 10 independent weightings, corresponding
to the 10 independent components of the metric.

With the lattice constructed in section \ref{sec-lattice},
one could average over all
Regge equations contained in a unit hypercube (it turns out that there are
65 of them in this particular case).
I define ${\cal R}_{avg}$ as
\begin{equation}
{\cal R}_{avg}(\delta) = \sum_{ij} \left( \frac {m_{ij}}{ {\delta}^2}
 \frac {\partial l_{ij}^2}{ \partial g_{\mu \nu} }
\right) {\cal R}_{ij}(\delta)
\label{eq:ravgdef}
\end{equation}
where the sum is over all legs (ij) contained in a unit hypercube,
and $m_{ij}$ is the combinatorial factor
\begin{equation}
m_{ij} = \frac {\mbox{number of 4-simplices inside unit hypercube that
have (ij) as a leg}}
{\mbox{total number of 4-simplices in the simplicial manifold that
have (ij) as a leg}}
\end{equation}
The  $1/{{\delta}^2}$ in the weighting is to keep the weighting factor
of order unity.  There are ten independent weightings, each
one corresponding to an independent component of the metric.
Plugging in the fifth order accurate expressions
for squared geodesic distances (\ref{eq:secondl})
into the Regge equations and
expanding ${\cal R}_{avg}(\delta)$ as a power series about $\delta = 0$ reveals
that the second order term of ${\cal R}_{avg}(\delta)$
vanishes when the Einstein equations are imposed at ${\cal O}$.
\begin{equation}
{( \frac { d^2 {\cal R}_{avg}}{d {\delta}^2} )}_{\delta
= 0} = 0
\label{eq:ravg2}
\end{equation}
This result holds for each of the ten independent weightings that
correspond to variations of the ten independent metric components.

One might wonder whether or not this result is an artifact of the
usage of Riemann normal coordinates.  To check, one can make
an arbitrary coordinate transformation on equation (\ref{eq:secondl})
before substituting the $l^2_{ij}$'s into the Regge equations.
It turns out that the result, eq. (\ref{eq:ravg2}), is
independent of the coordinate transformation, thus, the result
does not depend on the usage of Riemann normal coordinates.

\section{Numerical Results}
\label{sec-numerical}

In this section, I study the convergence properties of the
Regge equations numerically.  Instead of finding a power expansion for
squared geodesic distances (as in the previous section), I numerically
solve the geodesic equation
for different analytic solutions of the
vacuum Einstein equations (using the relaxation methods found in \cite{NRC}).
I then evaluate the Regge equations on
simplicial manifolds using these values.  The first non-zero order
term in a power expansion of ${\cal R}_{ij}(\delta)$ or
${\cal R}_{avg}(\delta)$ can
be found as the slope of the line of
a $ln({\cal R}(\delta))$ vs. $ln(1/ \delta)$ plot.  In all of the plots,
the error bars are actually smaller than the data points appearing on the
graph.  This fact is a result of the accuracy of the relaxation methods
used to solve the geodesic equations.  I have verified this by increasing
the resolution of the grid points by more than one order of magnitude, and
observing no change (to one part in $10^9$) in the values for the
computed geodesic
distances.

Instead of picking on one Regge equation to evaluate and plot, I actually plot
the sum of the absolute values of the Regge equations associated with the
links in a unit hypercube (summing over
the same links as in equation (\ref{eq:ravgdef})).
I define ${\cal R}_{abs}$ to be
\begin{equation}
{\cal R}_{abs} = \sum_{ij} \left|{{\cal R}_{ij}}\right|
\label{eq:rabsdef}
\end{equation}
I also plot the weighted average of the Regge
equations ${\cal R}_{avg}$ defined by
equation (\ref{eq:ravgdef}).

The results from section \ref{sec-analytic}
predict that ${\cal R}_{abs}$ should converge
like ${\delta}^2$.  Also, in section \ref{sec-analytic} it was shown that
the zeroth, first, and second order terms of ${\cal R}_{avg}$ vanish.  This
leaves on open question:
What is the first non-vanishing term of ${\cal R}_{avg}(\delta)$?

\subsection{Schwarzschild Metric}

The metric of the Schwarzschild solution
has the form
\begin{equation}
ds^2 = - \left( 1 - \frac {2 M}{r} \right) dt^2 +
{\left( 1 - \frac {2 M}{r} \right)}^{-1} dr^2  + r^2 {d \theta}^2
+ r^2 {\sin}^2\theta \, {d \phi}^2
\end{equation}
I set $M = 1$, and put the origin of the lattice just outside the event
horizon: $x^{\mu}_{\cal O} = (t_{\cal O},r_{\cal O},{\theta}_{\cal O},
{\phi}_{\cal O}) = (0,3,\pi/2,\pi)$.
The coordinate lattice spacing is $\Delta x^{\mu} = (\Delta t,\Delta r,
\Delta \theta, \Delta \phi)
= (\delta ,\delta / 2,
\delta / 4,\delta / 4)$.
The plot of ${\cal R}_{abs}$ is shown in Figure 3. The plot of
${\cal R}_{avg}$, with a weighting induced by the variation of the
time-time component of the continuum metric (the other weightings give
the same results),
is shown in Figure 4.

The figures show us a surprise.  First, we would expect
${\cal R}_{abs}$ to converge as
the second power in $\delta$, but Figure 3 shows a fourth order convergence.
Also, note the fourth order convergence of ${\cal R}_{avg}$.
To check these results,
I did the same calculation, using Kruskal coordinates, and moving ${\cal O}$
inside the event horizon.  In this case, the convergence properties
are shown to
be exactly the same as in Figures 3 and 4.

Something to notice about both
figures is the fact
that, even though the lengths of the geodesics are of the order of the
length scale defined by the curvature of the problem, the points corresponding
to $\delta = 1$ are already on the converging line.  This suggests that
Regge calculus may be superior to finite differencing in tracking the
geometry of black hole spacetimes.

\subsection{Kerr Metric}

The Kerr metric, which is the vacuum solution corresponding to
a rotating black hole, has the form
\begin{eqnarray}
ds^2 & = & - \left( \frac {r^2 + A^2 {\cos}^2\theta - 2 M r}{r^2 + A^2 {\cos}^2
\theta} \right) dt^2 -
\left( \frac {4 M A r {\sin}^2 \theta }{r^2 + A^2 {\cos}^2 \theta} \right)
dt \, d\phi +
\left( \frac {r^2 + A^2 {\cos}^2 \theta}{r^2 + A^2 - 2 M r} \right)
dr^2 \nonumber \\
 & &  + (r^2 + A^2 {\cos}^2 \theta) d{\theta}^2 +
\left( \frac { {(r^2 + A^2)}^2 - (r^2 + A^2 - 2 M r) A^2 {\sin}^2 \theta}
             {r^2 + A^2 {\cos}^2 \theta} \right) {\sin}^2 \theta \,
d{\phi}^2
\end{eqnarray}
where $M$ is the mass of the black hole and $A$ is the angular momentum
of the black hole divided by the mass: $A = J/M$.
I set $M = 1$, $A = 1/2$ (so that the spacetime is asymptotically predictable),
and put the origin of the lattice at
 $x^{\mu}_{\cal O} = (t_{\cal O},r_{\cal O},{\theta}_{\cal O},
{\phi}_{\cal O}) = (0,5,\pi/2,\pi)$.
The coordinate lattice spacing is $\Delta x^{\mu} = (\Delta t,\Delta r,
\Delta \theta, \Delta \phi)
= (1.5 \delta ,\delta ,
0.2 \delta ,0.3 \delta )$.
The plot of ${\cal R}_{abs}$ is shown in Figure 5, while the plot of
${\cal R}_{avg}$, with a weighting induced by the variation of the
time-time component of the continuum metric (the other weightings give
the same results),
is shown in Figure 6.

As can be seen, this example shows ${\cal R}_{abs}$ converging as the
second power of $\delta$, while ${\cal R}_{avg}$ still converges
as the fourth power of $\delta$.

\subsection{Kasner Metric}

The Kasner metric, which corresponds to a spatially homogeneous anisotropic
universe, has the form
\begin{equation}
ds^2 = - dt^2 + t^{2 p_1} dx^2 + t^{2 p_2} dy^2 + t^{2 p_3} dz^2
\end{equation}
where the constants $p_i$ satisfy $p_1 + p_2 + p_3 = 1$ and
$p_1^2 + p_2^2 + p_3^2 = 1$.  I set $p_1 = p_2 = 2/3$, and
$p_3 = -1/3$,
and put the origin of the lattice at
 $x^{\mu}_{\cal O} = (t_{\cal O},x_{\cal O},y_{\cal O},
z_{\cal O}) = (2,0,0,0)$.
The coordinate lattice spacing is $\Delta x^{\mu} = (\Delta t,\Delta x,
\Delta y, \Delta z)
= (\delta ,\delta ,
 \delta , \delta )$.
The plot of ${\cal R}_{abs}$ is shown in Figure 7, while the plot of
${\cal R}_{avg}$ , with a weighting induced by the variation of the
time-time component of the continuum metric (the other weightings
give similar results),
is shown in Figure 8.

We see that ${\cal R}_{abs}$ converges like the second power of $\delta$ and
${\cal R}_{avg}$ converges like the fourth power of $\delta$. These are the
same convergence properties found with the Kerr metric.

\section{Conclusions}

Both analytically and numerically, we see that the Regge equations
individually converge as the second power of the lattice spacing,
although this result is independent of whether or not the continuum
metric satisfies the Einstein equations.  We see that the second
order term of an average over local Regge equations vanishes
only when the continuum metric satisfies the Einstein
equations.  Numerical studies (Figures 4, 6, and 8) show that, for
certain analytic solutions of the Einstein equations,
the third order term of ${\cal R}_{avg}$ also vanishes.
Thus, Regge
calculus can be used as a fourth order method in numerical
relativity.

\section*{Acknowledgments}

I thank the members of the Syracuse Relativity Group for useful discussions
and suggestions. I particularly want to thank Rafael Sorkin,
without whose suggestions and ideas, the
results of this paper would not be nearly as strong.  I also want to thank
the Northeast Parallel Architecture Center (NPAC) for the use of their
computer facilities. This work was partially supported by a Fellowship
from Syracuse University and NSF ASC 93 18152 / PHY 93 18152 (ARPA
supplemented).

\end{document}